\documentstyle[twoside,fleqn,espcrc2,psfig]{article}

\newcommand{\BE}{\begin{equation}}
\newcommand{\EE}{\end{equation}}
\begin{document}

\title{Spins coupled to a $Z_2$-Regge lattice in 4d}

\author{E.~Bittner\address{Institut f\"ur Theoretische Physik, Universit\"at Leipzig, D-04109 Leipzig, Germany}$^,$\address{Atominstitut, Technische Universit\"at Wien, A-1040 Vienna, Austria}$^,$\thanks{E.B. was supported by the Fonds zur F\"orderung der wissenschaftlichen Forschung under project P14435-TPH and by the DFG through the Graduiertenkolleg ``Quantenfeldtheorie''.},
W.~Janke$^{\mbox{\scriptsize a}}$ and
H.~Markum$^{\mbox{\scriptsize b}}$
}

\begin{abstract}
We study an Ising spin system coupled to a fluctuating four-dimensional
$Z_2$-Regge lattice and compare with the results of the four-dimensional
Ising model on a regular lattice. Particular emphasis is placed on the
phase transition of the spin system and the associated critical exponents.
We present results from finite-size scaling analyses of extensive Monte Carlo
simulations 
which are consistent with mean-field predictions.
\end{abstract}

\maketitle

\noindent
\section{INTRODUCTION}
Spin systems coupled to fluctuating manifolds are studied as
a simple example for matter fields coupled to Euclidean quantum gravity. 
To describe the gravity sector we used 
%
the Discrete Regge Model~\cite{homo} which is both structurally and 
computationally much simpler than the Standard Regge Calculus with continuous 
link lengths.
Here numerical simulations can be done more efficiently by implementing look-up
tables and using the heat-bath algorithm. In the actual computations we took
the squared link lengths as
$q_{ij} \equiv q_l = b_l ( 1+\epsilon \sigma_l)$ with $\sigma_l = \pm 1$ and
$\epsilon = 0.0875$. Because a four-dimensional Regge
skeleton with equilateral simplices cannot be embedded in flat space,
$b_l$ takes different values depending on the type of the edge $l$. In particular
$b_l=1,2,3, 4$ for edges, face diagonals, body diagonals, and the hyperbody
diagonal of a hypercube. 

\section{MODEL AND OBSERVABLES}
We investigated the partition function
\begin{equation}\label{z}
Z=\sum_{\{s\}}\int D[q] \exp[-I(q) - K E(q,s)],
\end{equation}
where $I(q)$ is the gravitational action,
\begin{equation}
I(q) = - \beta_g \sum_t A_t \delta_t + \lambda \sum_i V_i.
\end{equation}
The first sum runs over all products of triangle areas $A_t$ times corresponding
deficit angles $\delta_t$ weighted by the gravitational coupling $\beta_g$.
The second sum extends over the volumes $V_i$ of the
4-simplices of the lattice and allows together with the cosmological constant
$\lambda$ to set an overall scale in the action.
The energy of  Ising spins $s_i  \in Z_2$,
\begin{equation}
E(q,s) = \frac{1}{2} \sum_{\langle ij \rangle} A_{ij}\frac{(s_i-s_j)^2}{q_{ij}},
\end{equation}
is defined as in two dimensions~\cite{physA}, with the barycentric area $A_{ij}$ associated
with a link $l_{ij}$, $A_{ij} = \sum _{\mbox{\scriptsize $t \supset l_{ij}$}} \frac{1}{3} A_t$.
We chose the simple uniform measure as in the pure gravity simulations~\cite{homo},
$D[q]=\prod_l dq_l {\cal F}(q_l)$.
The function ${\cal F}$ ensures that only
Euclidean link configurations are taken into account.

For every Monte Carlo simulation run we recorded the time series of the energy density
 $e=E/N_0$ and the magnetization density $m= \sum_i s_i /N_0$, with the lattice size $N_0=L^4$.
To obtain results for the various observables ${\cal O}$ at values of the spin coupling $K$
 in an interval around the simulation point $K_0$, we applied standard
reweighting techniques~\cite{ferrenberg}.

With the help of the time series we compute the specific heat,
$C(K)=K^2 N_0 (\langle e^2 \rangle- \langle e\rangle^2)$,
the (finite lattice) susceptibility,
$\chi(K)=N_0(\langle m^2 \rangle -\langle |m| \rangle^2)$,
the Binder parameter,
$U_L(K) = 1-\langle m^4 \rangle/3 \langle m^2\rangle^2$,
and various derivatives of the magnetization,
$d \langle |m| \rangle/dK$,
$d$ln$\langle |m| \rangle
/dK$, and
$d$ln$\langle m^2 \rangle/dK$.
All these quantities exhibit in the infinite-volume limit singularities at
$K_c$ which are shifted and rounded in finite systems.

\noindent
\section{SIMULATION RESULTS} 
In four dimensions it is generally accepted that the critical properties of
the Ising model on a static lattice are given by mean-field theory,
with logarithmic corrections. The finite-size formulas can be 
written as~\cite{ralph}
\begin{eqnarray}
\xi &\propto& L(\log L)^{\frac{1}{4}}, \label{eq:xi} \\
\chi  &\propto&  (L (\log L)^{\frac{1}{4}})^{\gamma/\nu}, \label{eq:chi2}\\
K_c(\infty)-K_c(L) &\propto& (L (\log L)^{\frac{1}{12}})^{-1/ \nu}, \label{eq:kc2}
\end{eqnarray}
where the critical exponents of mean-field theory are $\alpha=0$, $\beta=1/2$,
$\gamma=1$, and $\nu=1/2$.

The gravitational degrees of freedom of the 
partition function (\ref{z}) were updated with the heat-bath algorithm.
For the Ising spins we employed the single-cluster algorithm~\cite{wolff}.
Between measurements we performed $n=10$ Monte Carlo steps consisting of one
lattice sweep to update the squared link lengths~$q_{ij}$
followed by two single-cluster flips to
update the spins~$s_i$.

The simulations were done
for 
cosmological constant $\lambda = 0$ and gravitational coupling
$\beta_g=-4.665$. This $\beta_g$-value corresponds to a 
phase transition of the pure Discrete Regge Model~\cite{homo}.
The lattice topology is given by triangulated tori of size $N_0=L^4$ with
$L = 3$ up to 10.
From short test runs we estimated the location of the phase
transition of the spin model and set the spin coupling $K_0 = 0.024 \approx K_c$ in the long runs.

After an initial equilibration time we took about 
$100\,000$ measurements for each lattice size. Analyzing the time series we found integrated
autocorrelation times for the energy and the magnetization in the range of unity for all
lattice sizes. The statistical errors were obtained by the standard Jack-knife method
using 50 blocks.

Applying the reweighting technique we first determined the maxima of $C$,
$\chi$,  $d\langle |m| \rangle /dK$, $d$ln$\langle |m| \rangle /dK$,
and $d$ln$\langle m^2 \rangle/dK$. The locations of the maxima provide us with
five sequences of pseudo-transition points $K_{\rm max}(L)$ for which 
the scaling variable
$x=(K_{\rm max}(L) - K_c) (L (\log L)^\frac{1}{12})^\frac{1}{\nu}$ 
should be constant.
Using this fact we then have several 
possibilities to extract the critical exponent $\nu$ from (linear)
 least-square fits of the FSS ansatz with logarithmic
corrections~(\ref{eq:kc2}), 
\begin{eqnarray}
dU_L/dK &\cong&  (L (\log L)^\frac{1}{12})^{1/ \nu} f_0(x), \\
d{\rm ln}\langle |m|^p\rangle /dK &\cong&  (L (\log L)^\frac{1}{12})^{1/\nu} f_p(x),
\end{eqnarray}
to the data at the various $K_{\rm max}(L)$ sequences. 
We also performed fits of a naive power-law FSS ansatz.
The exponents $1/\nu$ resulting from fits 
using the data $L=4-10$ are collected in Table~\ref{tab:1}.
$Q$ denotes the standard goodness-of-fit parameter.
For our simulations all exponent estimates with the logarithmic corrections
and consequently also their weighted average $1/\nu = 2.028(7)$ 
are in agreement with the mean-field value $1/\nu=2$. 
With the naive power-law ansatz one also gets an estimate for $1/\nu$ close to
the mean-field value, but clearly separated from it.

\begin{figure}[t]
\psrotatefirst
\centerline{\hbox{
\psfig{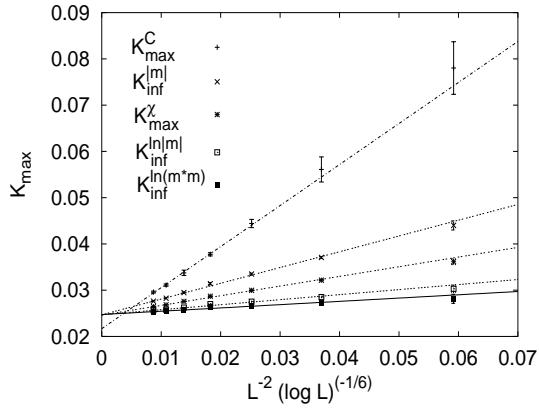}
}}
\vspace{-.6cm}
\caption{FSS extrapolations of pseudo-transition points $K_{\rm max}$ vs.
$(L (\log L)^{\frac{1}{12}})^{-1/\nu}$, assuming $\nu=0.5$. The error-weighted average of
extrapolations to infinite size yields $K_c=0.02464(4)$.}
\label{kc4diz2}
\vspace{-.8cm}
\end{figure}

\begin{table}[b]
\vspace{-.6cm}
\newlength{\digitwidth} \settowidth{\digitwidth}{\rm 0}
\catcode`?=\active \def?{\kern\digitwidth}
\caption[a]{Fit results for  $1/\nu$ in the range $L=4-10$ with a power-law ansatz with
logarithmic corrections.\label{tab:1}}
\begin{tabular}{llc}
\hline
\multicolumn{1}{c}{fit type}&
\multicolumn{1}{c}{$1/\nu$}&
\multicolumn{1}{c}{$Q$} \\ \hline\\[-0.3cm]
$dU/dK$ at $K^C_{\rm max}$ &~1.980(17)& 0.70\\
$d$ln$\langle |m|\rangle /dK$ at $K^{\ln\langle |m|\rangle}_{\rm inf}$ &~2.032(10)&0.59 \\
$d$ln$\langle m^2\rangle/dK$ at $K^{\ln\langle m^2\rangle}_{\rm inf}$  &~2.038(10)&0.55 \\
weighted average & ~2.028(7)&\\ \hline
$dU/dK$ at $K_c$  &~1.981(17)[13]&0.70 \\
$d$ln$\langle |m|\rangle /dK$ at $K_c$  &~2.027(9)[2]&0.95\\
$d$ln$\langle m^2\rangle/dK$ at $K_c$  &~2.034(9)[2]& 0.85\\
weighted average & ~2.025(6)&\\ \hline
\hline
overall average &  ~2.026(5)&\\ \hline
\end{tabular}
\end{table}
Assuming therefore $\nu=0.5$ we can obtain estimates for $K_c$ from linear least-square fits to 
the scaling behavior of the various $K_{\rm max}$ sequences, as shown in Fig.~\ref{kc4diz2}.
Using the fits with $L \ge 4$, the combined estimate from 
the five sequences leads to $K_c=0.02464(4)$.

Knowing the critical coupling we may reconfirm our estimates of $1/\nu$ by
evaluating the above quantities at $K_c$. As can be inspected in Table~\ref{tab:1},
the statistical errors of the FSS fits at $K_c$ are similar to those using the $K_{\rm max}$ 
sequences.  However, here we have to take into account the
uncertainty in our estimate of $K_c$. This error is computed by repeating the
fits at $K_c \pm \Delta K_c$ and indicated in Table~\ref{tab:1}
by the numbers in square brackets. In the computation of the weighted average we assume the
two types of errors to be independent.
As a result of this combined analysis we obtain strong evidence that the exponent 
$\nu$ agrees with the mean-field value of $\nu=1/2$.

\begin{figure}[t]
\centerline{\hbox{
\psfig{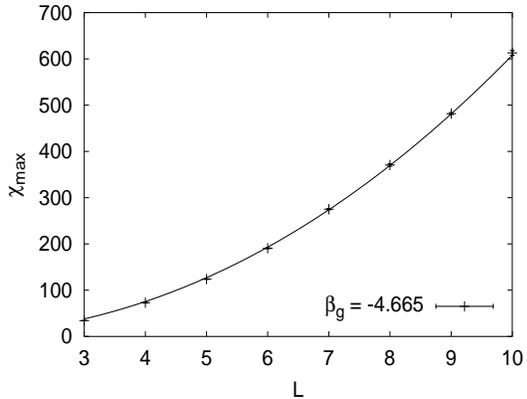}
}}
\vspace{-.6cm}
\caption{FSS of the susceptibility maxima $\chi_{\rm max}$. The exponent
entering the curve is set to the mean-field value $\gamma / \nu = 2$ for
regular static lattices.}
\vspace{-.8cm}
\label{fig_sus4diz2}
\end{figure}

To extract the critical exponent ratio $\gamma/\nu$ 
we use the scaling~(\ref{eq:chi2})
of the susceptibility $\chi$ at its maximum
as well as 
at $K_c$, yielding in the range $L=4 - 10$
 estimates of $\gamma/\nu=2.039(9)$ ($Q=0.42$) and
$\gamma/\nu=2.036(7)[4]$ ($Q=0.85$), respectively. 
These estimates for $\gamma/\nu$ are consistent with the mean-field value of 
$\gamma/\nu=2$.
In Fig.~\ref{fig_sus4diz2} this is demonstrated graphically
by comparing the scaling of $\chi_{\rm max}$ with a constrained one-parameter fit of 
the form $\chi_{\rm max} = c (L (\log L)^{\frac{1}{4}})^{2}$ with 
$c=4.006(10)$ ($Q=0.17$, $L \ge 6$).

%
%
\noindent
\section{CONCLUSIONS} 
We have performed a study of the Ising model coupled to fluctuating manifolds via
Regge Calculus. Analyzing the Discrete Regge Model with two permissible edge lengths it turns out
that the Ising transition shows the expected
 logarithmic corrections to the mean-field theory.
We have also studied the pure Ising model on a rigid lattice without presenting the results
in this short note.
The critical exponents of the phase transition of the Ising spins on a static lattice as
well as on a discrete Regge skeleton are both consistent with the exponents of
mean-field theory, $\alpha=0$, $\beta=1/2$, $\gamma=1$,
and $\nu=1/2$.
In summary, from our comparative analysis with uniform computer codes we 
conclude
that the phase transition of the Ising spin model coupled to
a discrete Regge skeleton exhibits the same critical exponents and the same
logarithmic corrections as on a static lattice.

\vspace{-.4cm}


\parskip1.2ex
\begin{thebibliography}{99}
%
\bibitem{homo}
W. Beirl, A. Hauke, P. Homolka, B. Krishnan, H. Kr\"oger, H. Markum,
and J. Riedler, Nucl. Phys.  B (Proc. Suppl.) 47 (1996) 625;
 W. Beirl, A. Hauke, P. Homolka, H. Markum, and J. Riedler,
Nucl. Phys. B (Proc. Suppl.) 53 (1997) 735;
J. Riedler, W. Beirl, E. Bitt\-ner, A. Hauke, P. Homolka, and H. Markum,
Class. Quant. Grav. 16 (1999) 1163.
%
\bibitem{physA}
E. Bitt\-ner, W. Janke, H. Markum, and J. Riedler, Physica A277 (2000) 204.
%
\bibitem{ferrenberg}
A.M. Ferrenberg and R.H. Swendsen, Phys. Rev. Lett. 61 (1988) 2635.
%
\bibitem{ralph}
R. Kenna and C.B. Lang, Phys. Lett. B264 (1991) 396;
Nucl. Phys. B393 (1993) 461.
%
\bibitem{wolff}
U. Wolff, Phys. Rev. Lett. 62 (1989) 361;
Nucl. Phys. B322 (1989) 759.
%
\end{thebibliography}
\end{document}